# Validating the Orbital Periods of the Coolest TESS Planet Candidates

Dillon J. Bass[1] and Daniel C. Fabrycky[1,2]
[1] Department of Astronomy & Astrophysics, University of Chicago, Chicago IL, 60637, USA
[2] National Science Foundation, 2415 Eisenhower Avenue, Alexandria VA, 22314, USA

## Abstract

When an exoplanet passes in front of its host star, the resulting eclipse causes an observable decrease in stellar flux, and when multiple such transits are detected, the orbital period of the exoplanet can be determined. Over the past seven years NASA's Transiting Exoplanet Survey Satellite (TESS) has discovered thousands of potential planets by this method, mostly with short orbital periods, although some have longer reported values over 100 days. These long orbital periods, however, are not easy to confirm due to frequent lengthy data gaps. Here we show that while many of these long period candidates likely have periods much shorter than reported, there are some TESS candidates with long periods to be found in the data. These candidates generally only have two reported transits, but the periods of duo-transits like this, and even candidates with three or more transits can be confirmed if the data rules out all possible shorter period aliases. Using TESS data, we confirm long orbital periods for nine candidate planets, and present five others that are likely long period ($P > 100$ days). Due to their long periods, these planets will also have relatively cool equilibrium temperatures. We present these TESS Objects of Interest, along with a variety of small corrections to other TESS orbital periods and three planet candidates with possible transit timing variations, with the goal of refining the TESS data set and enabling future research with respect to cool transiting planets.

*Unified Astronomy Thesaurus concepts:* Exoplanets (498); Transit photometry (1709)

## 1. Introduction

Since its launch in 2018, NASA's Transiting Exoplanet Survey Satellite (TESS; G. R. Ricker et al. 2014) has amassed a huge amount of data on bright stars across nearly the entire sky and detected more than 7000 transiting planet candidates.[3] The vast majority of these candidates, called TESS Objects of Interest (TOIs), have relatively short orbital periods due to a transit detection bias that favors finding planets orbiting close to their hosts (J. N. Winn & D. C. Fabrycky 2015). Additionally, the TESS observing cycle, which spends only ∼27 days at a time on each sector of the sky, is not conducive to detecting and confirming long-period planets. This is especially true in comparison to a mission like Kepler (W. J. Borucki et al. 2010), which maintained a continuous multiyear baseline for a single field. Yet, though the Kepler mission has discovered more than 60 very long-period planets ($P > 200$ days) orbiting fainter stars, the vast majority of exoplanets it has found still have much shorter periods (S. E. Thompson et al. 2018). For TESS, this short period bias was particularly obvious early in the mission when most sectors had only been observed once or twice. Over six years and eighty sectors later, however, TESS has begun to report a larger number of planet candidates that appear to have much longer periods.

Such planets, with possible orbital periods on the order of hundreds of days, are of interest for a variety of reasons. Because of their longer periods, they will generally have cooler equilibrium temperatures, and are more likely to be in the habitable zone of their host stars. This makes them interesting targets for habitability studies such as the recent JWST investigation of biosignatures on Hycean worlds (N. Madhusudhan et al. 2023), as well as candidates in the search for exo-moons and exo-rings (J. W. Barnes & D. P. O'Brien 2002). However, due to the overall scarcity of long-period transiting planet candidates in current data sets, such low temperature planets remain rare compared to warmer planets with shorter orbital periods.

In this paper, we make use of the TOI catalog, which records information on all candidate exoplanets extracted from the mission data (N. M. Guerrero et al. 2021), to attempt to refine the orbital periods of reported long-period candidates. The catalog contains measured and inferred parameters for each candidate, including a reported orbital period, and all the original data and processed light curves used to obtain these numbers.

The process of accurately identifying long-period TOIs in TESS data is made challenging by the mission's cyclic observing pattern which often results in hundreds of days between observations of a certain star. TESS spends about 27 days on each sector, so it is possible that a planet with a true orbital period of as few as 14 days could transit only once in a given sector. If that planet then transited again in a different sector many orbits later, it would appear to have an orbital period many times its true period, simply dependent on when TESS returned to that area of the sky. This issue is especially prevalent in sectors with fewer observations, and means that a TOI's orbital period generally cannot be fully confirmed without collecting enough data to rule out all possible integer fractions (known as aliases) of the observed period.

The fact that it usually requires more than two highly time-separated transits in order to confirm candidate orbital parameters is acknowledged by the authors of the TOI catalog (N. M. Guerrero et al. 2021) and well known in the literature. Similarly, it is well established that as the TESS mission progresses, the orbital periods of an increasing number of long-period candidates will become resolvable (J. Villanueva Steven et al. 2019; B. F. Cooke et al. 2021). For example, the mission's return to the southern ecliptic hemisphere discussed in B. F. Cooke et al. (2019) revealed many more potential long-period mono- and duo-transit TOIs. For these candidates, there are multiple ways to resolve period aliases, with the most straightforward being to simply observe a given star until there is another transit as in S. Gill et al. (2020), where the authors manually follow-up a TESS mono-transit in order to confirm the orbital period. Other methods to confirm the periods of planets with very few transits include taking radial velocity data, using data from other missions such as CHEOPS to refine period aliases as in Z. Garai et al. (2024), and writing probabilistic algorithms such as MonoTools from H. P. Osborn et al. (2022).

---

[3] TOI Releases: https://tess.mit.edu/toi-releases/.

In general, though, it is necessary to rule out all aliases in order to confirm the reported values for TOIs with longer periods and, as discussed above, the possibility for this type of analysis on TESS data has risen with the increasing length of the mission. Recent TESS discoveries include a 260 days planet from P. A. Dalba et al. (2022), a 482 days planet from I. Mireles et al. (2023), and multiple even longer period candidates are discussed in this paper. So while on the surface TESS does not appear to be the ideal method for finding long-period planets, its capabilities are only improving with time.

In this work, we present an analysis of all TOIs up through Sector 62, released in April 2023, that have reported periods longer than 27 days, with the goal of confirming the orbital periods of reported long-period planet candidates from TESS. Using the publicly available set of TESS data as described in Section 2, we analyze 266 TOIs to confirm and update reported orbital period values. From this analysis, we present four sets of results relating to the reported long orbital periods of TESS candidates. First, in Section 3.1, we present eight TOIs for which our algorithm found transits at an alias of the given period, thus significantly refining their periods. Second, in Section 3.2, we give a set of six TOIs for which our analysis reveals a slightly more accurate period than the previously reported value by noting an incremental shift in transit timing that can be attributed to a linear change in the ephemeris. Third, in Section 3.3, we report nine planet candidates from TESS for which we can confirm long periods, along with five likely long-period candidates, three of which are longer than any currently confirmed TESS exoplanet. Finally, in Section 3.4, we list three TOIs with apparent transit timing variations that cannot be easily corrected with an updated period. Overall, using the well known capability of TESS to discover more and more long-period planets as its mission lifetime lengthens, we create a reference for further more detailed work on any of these TESS candidates and refine existing orbital period values in the data set where possible.

## 2. Transit Search Code

### 2.1. Analysis Rationale

The suite of transit search code (D. Bass 2024) written in Python for this project primarily performs basic analyses on period aliases for long-period TESS planet candidates. The code checks for possible transits at aliases down to the minimum possible period based on the orbital period given in the TOI catalog, as further described in Section 2.2.

This type of analysis for TESS specifically is motivated by the existence of a group of reported TOIs with periods around 700 days that likely have shorter periods. This group was expected from the early TESS observation schedule, which revisited many sectors from its first cycle roughly 700 days after their initial observations. It is well known that these candidates periods could be refined with follow-up observations (B. F. Cooke et al. 2021).

In addition, the relation between orbital period and transit duration further indicates that the majority of candidates around the 700 days mark are likely to have shorter periods than currently reported. In general, longer period planets tend to have longer transit durations, though this is dependent on density of the host star and other factors such as eccentricity (E. B. Ford et al. 2008) and the transit impact parameter. In the TESS data, we see that the candidates with 700 days periods often do not have transit durations as long as might be expected from the lengths of their reported periods.

So, because of the structure of the TESS observing schedule, and distribution of the currently reported values, it is expected that the periods of many long-period candidates will eventually be confirmed at shorter period aliases.

### 2.2. Alias Search Code Overview

The Python code (D. Bass 2024) for this project focuses on analyzing light curves for TOIs based on their reported orbital period, and aliases of that period. The code uses the Lightkurve package (Lightkurve Collaboration et al. 2018) to download all available reductions[4] of the TESS data for a specific TOI, determines whether transits have occurred at times corresponding to possible period aliases, and then reports an overview of information for the TOI. This process produces a diagnostic PDF and allows us to evaluate whether the reported orbital period is accurate.

We ran this code on 266 of the longest reported orbital period planet candidates in the TOI catalog, (all TOIs with reported periods greater than 27 days as of 2023 April) and produced PDFs for each candidate. The analysis worked from the available reductions of TESS data to produce approximate transit fits, measured parameters, and graphs for easy visual analysis as in Figure 1.

After downloading all available light curves for a given TOI, period alias transit times are calculated, and the data is queried at all possible transit times for the smallest fraction of the period requested and up. To assess all relevant period aliases, the algorithm checks the light curves for every node of the 20th order Farey sequence of the reported orbital period (J. A. Carter & E. Agol 2013), with limits in place such that the same nodes are not analyzed multiple times.

Next, a chi squared best fit is performed using a three-parameter model based on the given transit depth and transit duration, and a fitted midpoint time for each alias. The median depth around the fitted midpoint is reported as the transit depth. If the best fit timing and depth agree with the predicted values and match across all available reductions of the light curve, then there is likely a transit. There are many cases in the TESS data where anomalies in the data result in errors with this process, but the existence of multiple reductions means that a single bad reduction will not invalidate the entire transit. This system is not a comprehensive method for identifying transits but is effective for alias searching since we already expect there may be undiscovered transits at a particular time.

---

[4] Usually SPOC (J. M. Jenkins et al. 2016), TESS-SPOC (D. A. Caldwell et al. 2020), and QLP (C. X. Huang et al. 2020) pipelines, as in Figure 1.

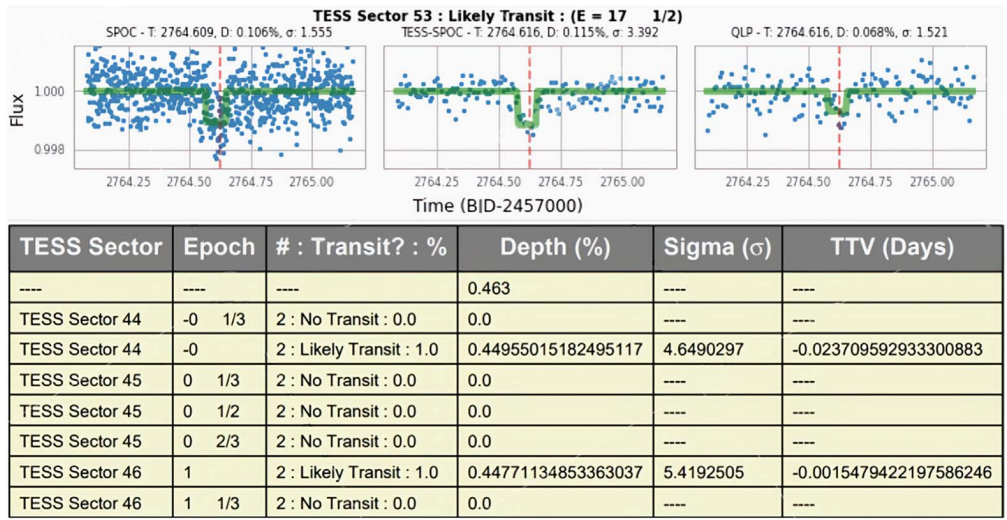

Figure 1. TOP: an example of a transit found and reported by the algorithm. The green line represents the best fit, and the dashed red line represents the predicted transit time for the alias. This specific case shows three reductions of the same data from different available pipelines. In each, a transit occurring 17.5 orbits after the originally reported transit is found. This is evidence that the reported period is at least double the true period for this particular planet. BOTTOM: an example of the table of transits produced by the algorithm. We report the sector searched, the alias of the period as the epoch after the original transit, a qualitative prediction of whether there was a transit, the measured depth, sigma heuristic, and calculated transit timing variation. These last three categories are only reported if the algorithm thinks there is a transit based on the data.

### 2.3. Analysis Products

The products of the analysis for each TOI are automatically collected into a single document showing relevant light curves, and a table of all transits that were searched. A PDF of the results for every analyzed TOI is available in the GitHub repository for this paper (D. Bass 2024). If a specific light curve is found to potentially have a transit, then it is presented as in the top panels of Figure 1. The entire list of all possible transits in the data is collected into a table as in the bottom panels of Figure 1. These results, taken with the reported transit parameters, are generally enough to make a determination for a given planet, classifying it into one of the result categories described in Section 3.

The overall purpose of the algorithm was to collect the data for each candidate into an easy to read format. Numerical results and associated uncertainties reported in Section 3 were recalculated for targets reported in this work after running the algorithm.

### 2.4. Sensitivity Analysis

In order to ensure that the transit detections reported in Section 3 were significant, we performed sensitivity testing for the algorithm on known transits in the TESS data. This is particularly useful for the results in Sections 3.1 and 3.3, which depend on the confidence with which we can detect that a light curve contains a transit. First, we found that for all known transits for the 266 long-period candidates used in this paper, the algorithm identified 1530 transits and missed 788. The approximately one third of transits that were missed generally had low transit depths and low signal to noise ratios. To quantify the algorithm's performance with respect to faint transit signals, we performed a more detailed sensitivity test as shown in Figure 2.

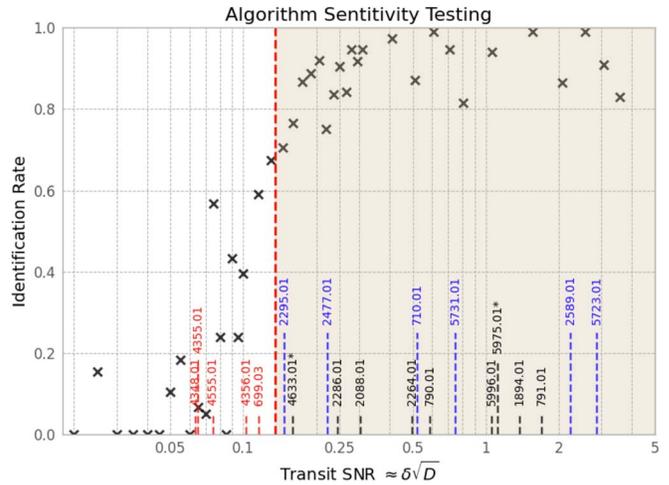

Figure 2. Each X marker in this figure represents transits from six similar signal strength TOIs, plotted at the mean signal value of the group, in order to ensure a large enough number of transits for every data point. The adopted signal strength threshold marked by the red dotted line is at $\delta\sqrt{D} = 0.135$. Along the horizontal axis, we plot the 20 candidates from Sections 3.1 (in blue) and 3.3 (in black or red) with red text indicating a low signal strength candidate. The asterisks denote candidates that are in both sections.

Assuming each transit is modeled by a simple boxcar function, we approximate the strength of the signal $S$ expected for these transits by taking the product of the transit depth $\delta$ and $\sqrt{D}$ where $D$ is the transit duration. This $S$ value is analogous to the signal to noise ratio formulae discussed in D. Kipping (2023) but dependent only on the transit depths and durations reported in the TOI catalog, and does not include the scatter of specific transits. Our sensitivity test shows that at the $\delta\sqrt{D} \approx 0.135$ limit in Figure 2, the identification rate is already 70%. Above this limit, the algorithm identifies around 90% of known transits successfully. On the bottom of Figure 2 we mark the TOIs reported in Section 3 to show for which candidates we can be confident that a transit on an alias was not missed. Candidates below the 0.135 signal limit used in this paper are noted as low signal targets, and discussed in more detail in Section 3.3.

**Table 1**
TOIs with Newly Discovered Period Aliases

| TOI # | $P_{\text{old}}$ | Alias | $P_{\text{new}}$ |
|---|---|---|---|
| 710.01 | 216.25 | 1/3 | 72.08 |
| 2295.01 | 60.07 | 1/2 | 30.04 |
| 2477.01 | 313.01 | 1/9 | 34.78 |
| 4356.01 | 579.64 | 1/2 | 289.82 |
| 4633.01 | 543.88 | 1/2 | 271.94 |
| 5723.01 | 170.65 | 1/3 | 56.88 |
| 5731.01 | 270.30 | 1/3 | 90.1 |
| 5975.01 | 373.04 | 1/2 | 186.52 |

**Table 2**
TOIs with Small Period Corrections

| TOI # | $P_{\text{old}}$ | $\Delta P$ (day) | $P_{\text{new}}$ |
|---|---|---|---|
| 588.01 | $39\overset{\text{d}}{.}4714 \pm 0.00001$ | +0.00046 | $39\overset{\text{d}}{.}4718 \pm 0.00002$ |
| 694.01 | $48\overset{\text{d}}{.}0512 \pm 0.00002$ | −0.00081 | $48\overset{\text{d}}{.}0503 \pm 0.00015$ |
| 1793.01 | $55\overset{\text{d}}{.}09 \pm 0.0199$ | −0.0077 | $55\overset{\text{d}}{.}0823 \pm 0.00001$ |
| 2264.01 | $121\overset{\text{d}}{.}3458 \pm 0.00041$ | −0.0072 | $121\overset{\text{d}}{.}353 \pm 0.001$ |
| 2286.01 | $179\overset{\text{d}}{.}412 \pm 0.0014$ | −0.0060 | $179\overset{\text{d}}{.}406 \pm 0.0013$ |
| 5575.01 | $32\overset{\text{d}}{.}07362 \pm 0.000001$ | −0.0009 | $32\overset{\text{d}}{.}0727 \pm 0.00007$ |

## 3. Results

### 3.1. Missed Period Alias Transits

We present a set of eight TOIs for which previously unreported transits were flagged by our algorithm at aliases of the reported orbital period. For example, if a planet was reported as having a 200 days period, but analysis by the algorithm found a transit 100 days after the original transit, then the true period can be refined by half, if not necessarily confirmed. The eight TOIs matching this category are detailed in Table 1. The new reported periods in the table are calculated by simple division of the old reported period by the newly found alias.

Because of the signal limit of our algorithm (the transits in this section are all above the limit from Figure 2) it seems likely that there are many remaining missed aliases below our sensitivity threshold.

Finally, while seven of these new aliases have not been previously reported, TOI-4633.01 has since been confirmed as a TOI-4633c with a 271 days period by N. L. Eisner et al. (2024).

### 3.2. Small Period Errors

We present six TOIs in Table 2 for which visual inspection noted deviation from current TESS ephemeris over multiple orbits that could be corrected by slightly adjusting the reported periods. A more complete analysis of period values could certainly produce many small corrections to other TOIs, we simply present a few obvious ones here.

Such small errors in the reported periods may be caused by using only a few initially discovered transits to calculate periods, and would likely be resolved by repeating the original period calculation with the most recent data. These errors, though small, build up over time, and the more orbits after the original data used to calculate the period, the worse transit predictions will become. For example, an error of 0.002 days will cause the ephemeris prediction to be off by about 90 minutes after 30 orbits. One reason this is potentially problematic is that such errors can be easily confused with transit timing variations caused by other factors such as additional planets in the system, and indeed any analysis of such variations would be incorrect if the reported period that calculates predicted ephemeris was incorrect.

To determine whether a candidate's ephemeris could be corrected with an improved period, we used the transit time calculated by the algorithm for each transit and the predicted ephemeris based on the transit time and period given in the TOI catalog. The difference between these two values (the TTV value in Figure 1) was then plotted for each available transit. By this method, we demonstrate that the period deviations for the candidates in the table are consistent from orbit to orbit, and therefore unlikely to be true transit timing variations (TTVs). If the raw transit mistiming increases or decreases linearly, this indicates that the deviation can be fixed by adjusting the period, whereas a nonlinear pattern of larger amplitude than the measurement error indicates a possibly real transit timing deviation.

Upon finding a candidate with a significantly sloped linear trend indicating an inconsistent period, we recalculated the orbital period. The new periods, listed in Table 2, were determined using the linear best fit of the transit timing data and the uncertainties on the new periods were calculated from the standard error on the slopes of the linear fits.

To further ensure that these corrections were significant, we only report candidates that had calculated period corrections much larger than the initially reported error in the period from the TOI catalog. As seen in the table, the $\Delta P$ values are much greater than the error previously reported in the catalog, and thus represent significant corrections. One exception to this is TOI-1793.01, which has been confirmed as HD 95338 b by M. R. Díaz et al. (2020). Although the corrected period remains consistent with previous estimates, our updated value significantly refines the period by reducing its uncertainty by more than an order of magnitude, so it is still included in the table.

### 3.3. Likely Long Period Planet Candidates

In the TESS data, we found fourteen different TOIs with periods that appear to truly be greater than 100 days, twelve of which have data at every alias. These candidates are listed in Table 3 along with their period, duration, and possible missing aliases. We believe nine of these have high enough signal to be confirmed with periods longer than 100 days. This is, as expected, much less than the total number of TOIs with reported periods over 100 days, but is still a significant number. The current longest period confirmed TESS planet is TOI-4600 c with a period of 482 days from I. Mireles et al. (2023), and five of these candidates may be even longer than that.

Though it seems likely that most of these candidates do in fact have long periods ($P > 100$), many remain uncertain for various reasons. TOI-1894.01, for example, is reported at 739 days, but there is no data at the 1/2 period alias. The 739 days period is by no means ruled out—the 19 hr transit duration is certainly long enough to support this period—but more observations are required to rule out the 370 days period alias. Occasionally, there are also transits that are simply missed by the pipelines, as was the case with TOI-4633.01: this

**Table 3**
Potential Long Period TOIs

| TOI # | Period (days) | Duration (hr) | Missing Aliases | Note |
|---|---|---|---|---|
| 699.03 | 672.64 | 7.49 | None | Low signal! |
| 790.01 | 199.58 | 11.70 | None | Very clear |
| 791.01 | 139.30 | 12.09 | None | Very clear |
| 1894.01 | 739.24 | 19.24 | 1/2 | Clear duo-transit |
| 2088.01 | 124.73 | 6.62 | None | Very clear |
| 2264.01 | 121.34 | 6.63 | None | Very clear |
| 2286.01 | 179.41 | 11.17 | None | Very clear |
| 4348.01 | 602.24 | 8.94 | None | Low signal! |
| 4355.01 | 674.23 | 12.23 | 1/2, 1/3 | Low signal! |
| 4356.01 | 289.83 | 4.57 | None | Low signal! |
| 4555.01 | 522.63 | 9.86 | None | Low signal! |
| 4633.01 | 271.94 | 11.13 | None | Down from 543 days |
| 5975.01 | 186.52 | 15.33 | None | Long duration |
| 5996.01 | 109.05 | 7.39 | None | Very clear |

**Note.** TOI-2088.01 was listed as a confirmed planet, TOI-2088 b, with a confirmed period of 124 days by A. S. Polanski et al. (2024).

candidate has a single light curve reduction at the $\frac{1}{2}$ period that shows a transit that matches the other two reported light curves. Others, like TOI-699.03 and 4348.01 technically have data covering all possible period aliases, but have high scatter in their light curves and signals below the limit discussed in Section 2.4, meaning any currently missed transits would likely be undetectable to our analysis.

Overall, out of the fourteen candidates, the seven with periods less than 200 days have enough data and high enough signal to be certain of their orbital periods. Each one has more than two transits in the data set, and some have 10 or more transits. These TOIs are marked with the note "very clear" in Table 3 to indicate that there is little uncertainty in their observations. TOI-4633.01 also has high enough signal for the period to be confirmed, after being updated with the newly discovered half-period alias, and has also been confirmed by N. L. Eisner et al. (2024) since the analysis done for this paper was completed. In addition, TOI-1894.01 is certainly a long-period candidate, even if it turns out to have half its currently reported period. So, it is these nine TOIs that we confirm to truly have long periods.

The other five candidates in the table have transit signals below the limit of our algorithm, are generally longer period, and have only two or three known transits. This makes it harder to be certain of their periods, and also harder to know if transits were missed. Thus, the five candidates noted in the table as "Low signal!" are too uncertain to confirm periods for in this work, even if all aliases are covered, as evidenced by the detection rates in Figure 2.

Apart from the candidates in Table 3, there are still many TOIs with long reported periods. These TOIs' periods, however, should be considered as maximum values, and not representative of true orbital periods. The same analysis that found the candidates that do appear to actually have long periods in Table 3 also found that almost every other candidate with a listed period greater than 500 days was a duo-transit with almost no data on period aliases. Coupled with their often very short transit durations, it is clearly necessary to obtain much more data to rule out aliases for the majority of reported long-period planet candidates in the TESS data.

### 3.4. Transit Timing Variations (TTVs)

In our analysis, we found three TOIs with transits that did not visually appear to exactly match their predicted transit times, similar to Section 3.2, but with no way to correct them simply by adjusting the period value, indicating that there were real transit timing variations present. As in Section 3.2, there are certainly many more real TTVs in the TESS data that could be discovered with a more thorough strategy. We simply report a few obvious ones. For example, in our analysis we quickly noticed TTVs for TOI-2525 b which were already confirmed by T. Trifonov et al. (2023). In that case, the authors were able to perform a more complete analysis resulting in the confirmation of two companion perturbing planets in the system. While it technically only takes three transits to confirm a TTV, more are usually needed in order to do such in depth work, and the candidates here have relatively few transits. These TOIs are, however, good possibilities for future observing and analysis due to their obvious TTVs, and their magnitudes which are generally much brighter than known TTV candidates from Kepler. These new TTVs, all of which have fairly large amplitudes, are presented in Table 4.

To be sure that these TTVs are significant, we compare the calculated transit time ($t_{\text{calc}}$, from the TESS catalog) with the measured one ($t_{\text{real}}$ calculated in this work) and their associated uncertainties in Table 4. The $t_{\text{calc}}$ error values are derived from the error on the given $t_0$ transit time reported in the TESS catalog and the number of orbits away from $t_0$ epoch. The uncertainty for the measured value $t_{\text{real}}$ was calculated using the chi squared best fit of the specific transit, similar to the method the initial algorithm used to detect transits.

The reported uncertainties for $t_{\text{real}}$ use the approximation that an increase of one in the chi squared value is a one sigma change in the transit time parameter for fixed depth and duration. So, transit times outside of the reported uncertainties on $t_{\text{real}}$ produce significantly worse chi squared fits. For each transit in the table, comparing these transit times and uncertainties indicates a difference of more than three sigma between the two. The sole exception is the $n = 11$ transit of TOI-5581, but this transit is still included since the candidate has two other much more significant TTVs and this one is still greater than 1 sigma itself.

### 4. Discussion

Compared to the current literature on known long-period and cool equilibrium temperature planets we present a relatively large number of candidates that, if confirmed, could significantly expand the data pool. Since the majority of TESS candidates have short orbital periods and orbit close to their host stars, they also generally have higher equilibrium temperatures. In comparison, the long-period candidates in this paper appear to have much cooler equilibrium temperatures. This is particularly true for the seven possible longest period candidates with periods greater than 200 days reported in Section 3.3. As seen in Figure 3, these candidates may be in the range of cool long-period candidates from Kepler, but have considerably brighter apparent magnitudes. These magnitudes arise systematically because TESS observes a much larger number of stars, but for shorter durations per star compared to Kepler. This characteristic may benefit follow-up observations from observatories that have difficulty targeting very faint objects.

Table 4
Possible New Large TTVs

| TOI # | Mag | n | $t_{calc}$ (days) | $t_{real}$ (days) | TTV (minute) | $\sigma$ |
|---|---|---|---|---|---|---|
| 5552.01 | 11.99 | −23 | 1824.191 ± 0.021 | 1824.412 ± 0.002 | 318 ± 2 | 10.5 |
| 5552.01 | 11.99 | 15 | 2912.678 ± 0.013 | 2912.747 ± 0.002 | 99 ± 2 | 5.2 |
| 5581.01 | 11.83 | −16 | 1854.367 ± 0.013 | 1854.4635 ± 0.0012 | 140 ± 2 | 7.4 |
| 5581.01 | 11.83 | −11 | 2022.053 ± 0.009 | 2022.1033 ± 0.0015 | 72 ± 2 | 5.5 |
| 5581.01 | 11.83 | 11 | 2759.873 ± 0.009 | 2759.8834 ± 0.0012 | 14 ± 2 | 1.1 |
| 5975.01 | 10.63 | −1.5 | 1849.070 ± 0.005 | 1848.991 ± 0.004 | 114 ± 5 | 12 |

**Note.** "$n$" refers to the $n$th transit after initial $t_0$ from the TOI catalog. Times refer to mid-transit TBJD. Magnitude is taken from TESS catalog. $t_{calc}$ refers to the transit time calculated from the catalog transit time and period, while $t_{real}$ refers to the new value measured in this work.

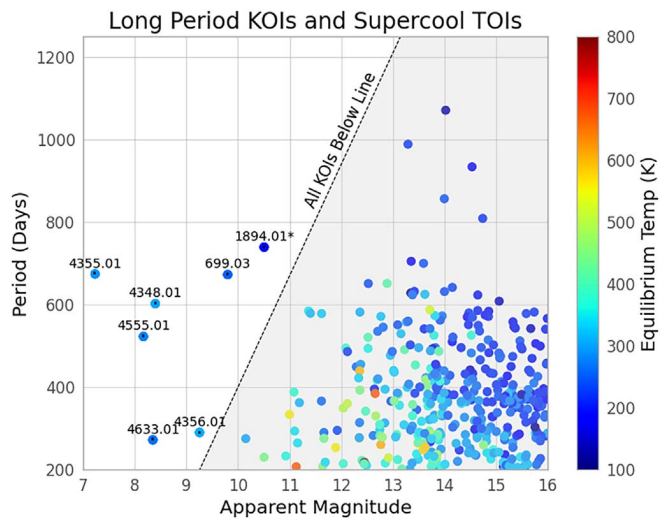

**Figure 3.** A comparison between the seven longest period (if fully confirmed at current values) candidates from this paper, and the KOI catalog (S. E. Thompson et al. 2018). Equilibrium temperatures and apparent magnitudes are taken directly from the respective TOI and KOI catalogs. Note that TOI-1894's period could be half what is reported.

Overall, we provide an initial resource for selecting likely candidates for future studies of long-period planets. We find that, in general, TESS candidates with long reported periods are duo-transits with almost no data in between their two reported transits. There are, however, a significant number of candidates which do have enough data to be confirmed as long period, as discussed in Section 3.3. These super-long-period cool equilibrium temperature candidates specifically, and the TTV possibilities in Section 3.4, are the most promising candidates for future research.

Additionally, it is sure that the alias data for these candidates will only become more complete as the TESS mission continues. So, once fully confirmed, multiple candidates (TOIs 699.03, 1894.01, 4348.01, 4355.01, and 4555.01) discussed in this paper have the potential to set the record for longest period TESS exoplanet to date, and the occurrence of such candidates in this data set reinforces the capability of TESS to find planet candidates with periods as long as many of the longest planets in the Kepler data set.

### Acknowledgments

We gratefully acknowledge the Hoeft Fund for Undergraduate Research for supporting D.J.B. This material is based upon work supported by (while serving at) National Science Foundation, for D.C.F.

### ORCID iDs

Daniel C. Fabrycky  https://orcid.org/0000-0003-3750-0183